\documentclass[a4paper,twoside,11pt]{article}
\usepackage{french,xspace}

\def\bbib{\begin{thebibliography}{999}}
\def\ebib{\end{thebibliography}}

\def\beq{\begin{equation}}
\def\eeq{\end{equation}}

\def\beqn{\begin{eqnarray}}
\def\eeqn{\end{eqnarray}}

\def\beqnn{\begin{eqnarray*}}
\def\eeqnn{\end{eqnarray*}}

\def\barr{\begin{array}}
\def\earr{\end{array}}

\def\bit{\begin{itemize}}
\def\eit{\end{itemize}}

\def\wh#1{\widehat{#1}}
\def\disp#1{{\displaystyle #1}}

\def\ER{\mbox{I\kern-.25em R}}
\def\EC{\mbox{C\kern-.8em C}}
\def\EZ{\mbox{Z\kern-.55em Z}}
\def\EN{\mbox{N\kern-.8em N}}

\def\bm#1{\mbox{\boldmath $#1$}}
\def\sb{{\bm s}}
\def\ub{{\bm u}}
\def\xb{{\bm x}}
\def\yb{{\bm y}}
\def\Ab{{\bm A}}
\def\Bb{{\bm B}}
\def\Fb{{\bm F}}
\def\Rb{{\bm R}}
\def\Sb{{\bm S}}
\def\Ub{{\bm U}}

\def\thetab{\bm{\theta}}
\def\lambdab{\bm{\lambda}}
\def\Lambdab{\bm{\Lambda}}
\def\Sigmab{\bm{\Sigma}}
\def\lra{\longrightarrow}
\def\intg{\int\kern-1.1em\int}
\def\intd{\int\kern-.8em\int}

\def\apriori{{\em a priori} }
\def\aposteriori{{\em a posteriori} }

\def\argmaxs#1#2{\mbox{arg}\max_{#1}\left\{{#2}\right\}}
\def\argmin#1#2{\mathop{\mbox{arg}\min}_{#1}\left\{{#2}\right\}}
\def\argmax#1#2{\mathop{\mbox{arg}\max}_{#1}\left\{{#2}\right\}}
\def\d#1{\,\mbox{d}#1}
\def\esp#1{\mbox{E}\left\{ #1 \right\}}
\def\espx#1#2{\mbox{E}_{#1}\left\{ #2 \right\}}
\def\expf#1{\exp\left[ {#1} \right]}

\def\ent#1{\hbox{H}\left[#1\right]}
\def\dent#1{\delta\hbox{H}\left[#1\right]}
\def\rent#1{\hbox{D}\left[#1\right]}
\def\kullback#1{\hbox{K}\left[#1\right]}
\def\pent#1{\hbox{PE}\left[#1\right]}
\def\mdiv#1{\hbox{J}\left[#1\right]}
\def\infmut#1{\hbox{I}\left[#1\right]}

\def\j{\mbox{j}}

\def\pxtheta{p(x | \thetab)}
\def\pxthetas{p(x | \thetab^*)}
\def\pxthetasd{p(x | \thetab^*+\Delta\thetab)}

\def\tr#1{\hbox{tr}\left(#1\right)}
\def\det#1{\hbox{d\'et}\left(#1\right)}

\title{Entropie en Traitement du Signal \\ Entropy in Signal Processing}

\author{Ali Mohammad-Djafari \\
Laboratoire des Signaux et Syst\`emes ({\sc cnrs-sup\'elec-ups}) \\ 
{\sc sup\'elec} \\ 
Plateau de Moulon, 91192 Gif-sur-Yvette Cedex, France.}

\date{~}

\begin{document}
\maketitle
\thispagestyle{empty}

\vspace*{-1cm}
{\small 
\section*{R\'esum\'e} 
Le principal objet de cette communication est de faire une r\'etro perspective 
succincte de l'utilisation de l'entropie et du principe du maximum d'entropie 
dans le domaine du traitement du signal. 
Apr\`es un bref rappel de quelques d\'efinitions et 
du principe du maximum d'entropie, nous verrons
successivement comment l'entropie est utilis\'ee en s\'eparation de sources, 
en mod\'elisation de signaux, en analyse spectrale et pour la r\'esolution 
des probl\`emes inverses lin\'eaires.  

\medskip\noindent{\bf Mots cl\'es :} 
Entropie, Entropie crois\'ee, Distance de Kullback, 
Information mutuelle, Estimation spectrale, Probl\`emes inverses

\section*{Abstract} 
The main object of this work is to give a brief overview of the different 
ways the entropy has been used in signal and image processing. 
After a short introduction of different quantities related to the entropy 
and the maximum entropy principle, we will study their use in different 
fields of signal processing such as: source separation, model order 
selection, spectral estimation and, finally, general linear inverse problems.  

\medskip\noindent{\bf Keywords :} 
Entropy, Relative entropy, Kullback distance, 
Mutual information, Spectral estimation, Inverse problems
}

\section{Introduction}
En 1945, Shannon \cite{Shannon48} a introduit la notion de l'entropie     
associ\'ee \`a une source qui est mod\'elis\'ee par une variable al\'atoire 
discr\`ete $X$, comme la moyenne de la quantit\'e d'information apport\'ee 
par les r\'ealisations de cette variable. 
Depuis cette date, cette notion a eu un tr\`es grand  
usage dans le domaine du traitement de l'information et particuli\`erement 
en codage et compression des donn\'ees en t\'el\'ecommunications. 

En 1957, Jaynes \cite{Jaynes57a,Jaynes57b,Jaynes68} a introduit le principe 
du maximum d'entropie pour l'attribution 
d'une loi de probabilit\'e \`a une variable al\'eatoire lorsque la  
connaissance sur cette variable est incompl\`ete. 

En 1959, Kullback \cite{Kullback59} a introduit une mesure de l'information 
relative (entropie relative) d'une loi de probabilit\'e par rapport 
\`a une autre. Cette mesure a aussi \'et\'e consid\'er\'ee comme une 
mesure de distance entre ces deux lois. 

Depuis, ces notions ont eu une influence importante et 
un usage \'etendu dans divers domaines du traitement de l'information, 
de l'inf\'erence en g\'en\'eral, mais aussi du traitement du signal 
et des images. 

Le principal objet de ce travail est de fournir une vue synth\'etique et 
br\`eve des principaux usages de ces notions en traitement du signal. 
Apr\`es un rappel de quelques d\'efinitions, des relations importantes 
entre les diff\'erentes quantit\'es et l'expos\'e du principe du maximum
d'entropie, nous verrons successivement comment l'entropie est utilis\'ee 
en s\'eparation de sources, en mod\'elisation de signaux, 
en analyse spectrale et pour la r\'esolution 
des probl\`emes inverses lin\'eaires. 

\subsection{Rappels et d\'efinitions}

L'entropie associ\'ee \`a une variable al\'eatoire scalaire discr\`ete 
$X$ avec des r\'ealisations $\{x_1,\cdots,x_N\}$ et la distribution de 
probabilit\'es $\{p_1,\cdots,p_N\}$ mesure son d\'esordre. Elle est 
d\'efinie par 
\beq
\ent{X}= - \sum_{i=1}^N  p_i \, \ln p_i. 
\eeq
Avec quelques pr\'ecautions, cette d\'efinition peut \^etre \'etendue 
au cas d'une variable al\'eatoire continue $X$ avec une densit\'e de
probabilit\'e $p(x)$ par 
\beq
\ent{X}= - \int p(x) \, \ln p(x) \d{x}.  
\eeq
Par extension, si on consid\`ere un couple de variables al\'eatoires 
$(X, \Theta)$ avec des lois 
$p(x)$, $p(\theta)$, $p(\theta | x)$, $p(x | \theta)$ et $p(x,\theta)$, 
on peut d\'efinir les entropies respectivement associ\'ees : 
\bit
\item Entropie de $\Theta$ : 
\beq
\ent{\Theta} = - \int p(\theta) \, \ln p(\theta) \d{\theta}
\eeq
\item Entropie de $\Theta$ conditionnellement \`a $X=x$ : 
\beq
\ent{\Theta | x} = - \int p(\theta | x) \, \ln p(\theta | x) \d{\theta}
\eeq
\item Entropie de $X$ conditionnellement \`a $\Theta=\theta$ : 
\beq
\ent{X | \theta} = - \int p(x | \theta) \, \ln p(x | \theta) \d{x}
\eeq
\item Entropie de $(X,\Theta)$ : 
\beq
\ent{X,\Theta} = - \intd p(x,\theta) \, \ln p(x,\theta) \d{x} \d{\theta}. 
\eeq
\eit
Avec ces d\'efinitions on d\'efinit aussi les quantit\'es 
suivantes : 
\bit
\item Diff\'erence entropique de $p_1$ et $p_2$ : 
\beq
\dent{p_1, p_2} = \ent{p_1} - \ent{p_2} 
\eeq
\item Entropie relative de $p_1$ par rapport \`a $p_2$ : 
\beq
\rent{p_1 : p_2} = - \int p_1(x) \, \ln \frac{p_1(x)}{p_2(x)} \d{x}
\eeq
\item Distance de Kullback de $p_1$ par rapport \`a $p_2$ : 
\beq
\kullback{p_1 : p_2} = -\rent{p_1 : p_2} 
= \int p_1(x) \, \ln \frac{p_1(x)}{p_2(x)} \d{x}
\eeq
\item Information mutuelle entre $\Theta$ et $X$ : 
\beqn
\infmut{\Theta, X} 
&=& \espx{X}{\dent{p(\theta), p(\theta | x)}} 
= \espx{\Theta}{\dent{p(x), p(x | \theta)}} 
\\ 
&=& \ent{X} - \ent{X | \Theta}= \ent{\Theta} - \ent{\Theta | X}
\eeqn
o\`u 
\beqn
\ent{\Theta | X} &=& \espx{X}{\ent{\Theta | x}} 
= \int \ent{\Theta | x} \, p(x) \d{x} 
\\ 
\ent{X | \Theta} &=& \espx{\Theta}{\ent{X | \theta}} 
= \int \ent{X | \theta} \, p(\theta) \d{\theta}  
\eeqn
avec les relations suivantes entre ces diff\'erentes quantit\'es : 
\beqn
\ent{X,\Theta} &=& \ent{X} + \ent{\Theta | X}  
= \ent{\Theta} + \ent{X | \Theta}  
= \ent{X} + \ent{\Theta} - \infmut{\Theta, X}
\\ 
\infmut{\Theta, X} 
&=& \rent{p(x,\theta) : p(x) p(\theta)}  
= \espx{X}{\rent{p(\theta | x) : p(\theta)}}  
= \espx{\Theta}{\rent{p(x | \theta) : p(x)}}.
\eeqn
\eit
On peut aussi remarquer les propri\'et\'es suivantes :
\bit
\item L'information mutuelle $\infmut{\Theta, X}$ 
est une fonction concave de $p(\theta)$ pour $p(x | \theta)$ fix\'ee et 
une fonction convexe de $p(x | \theta)$ pour $p(\theta)$ fix\'ee, et on a 
$\infmut{\Theta, X} \ge 0$ 
avec \'egalit\'e si $X$ et $\Theta$ sont ind\'ependantes. 
Cette propri\'et\'e est utilis\'ee en communication pour d\'efinir 
la capacit\'e d'un canal lorsque $X$ est transmis et 
$\Theta$ est re\c cu :
\beq
C = \argmaxs{p(\theta)}{\infmut{\Theta, X}}
\eeq

\item L'entropie relative $\rent{p_1 : p_2}$ est invariante par 
changement d'\'echelle mais n'est pas sym\'etrique. 
C'est pourquoi on introduit  
\beq
\mdiv{p_1,p_2} = \rent{p_1 : p_2} + \rent{p_2 : p_1}, 
\eeq
qui est sym\'etrique et invariante par changement d'\'echelle,  
comme une mesure de divergence entre $p_1(x)$ et $p_2(x)$. 

\item La puissance entropique (PE) d'une loi $p(x)$ est d\'efinie 
comme la variance d'une loi gaussienne ayant la m\^eme entropie. 
En notant que l'entropie d'une loi gaussienne est 
$\frac{1}{2} \, \ln (2\pi e \sigma^2)$, 
on obtient 
\beq
\pent{p} 
= \expf{2 \, [\ent{p}-\frac{1}{2} \, \ln (2\pi e)]} 
= \expf{2 \, \dent{p,{\cal N}(0,1)}}. 
\eeq

$\pent{p}$ est une mesure de proximit\'e de $p$ \`a une densit\'e 
gaussienne r\'eduite. 
\eit

\subsection{Lien entre entropie et vraisemblance}

Consid\'erons le probl\`eme de l'estimation des param\`etres $\thetab$ 
d'une loi de probabilit\'e $\pxtheta$ \`a partir d'un $n$-\'echantillon  
$\xb=\{x_1,\cdots,x_n\}$. 
La log-vraisemblance de $\thetab$ est d\'efinie par 
\beq \label{logv}
L(\thetab)=\sum_{i=1}^n \, \ln \pxtheta. 
\eeq
Maximiser $L(\thetab)$ par rapport \`a $\thetab$ donne l'estimation au 
sens du maximum du vraisemblance (MV). 
Notons que $L(\thetab)$ d\'epend de $n$, c'est 
pourquoi on peut s'int\'eresser \`a $\frac{1}{n} L(\thetab)$ et d\'efinir 
\beq 
\bar{L}(\thetab)=\lim_{n\mapsto\infty} \frac{1}{n} L(\thetab) 
= \esp{\, \ln \pxtheta} = \int \pxthetas \, \ln \pxtheta \d{x}, 
\eeq
o\`u $\thetab^*$ est le pr\'esum\'e ``vrai'' vecteur des param\`etres et 
$\pxthetas$ la loi de probabilit\'e correspondante. 
On peut alors noter que 
\beq 
\rent{\pxthetas : \pxtheta}
=-\int \pxthetas \, \ln \frac{\pxtheta}{\pxthetas} \d{x} 
=\int \pxthetas \, \ln \pxthetas \d{x} + \bar{L}(\thetab)
\eeq
et que 
\[
\argmax{\thetab}{\rent{\pxthetas : \pxtheta}}
=\argmax{\thetab}{\bar{L}(\thetab)}.
\]

\subsection{Lien entre entropie et la matrice d'information de Fisher} 
 
On consid\`ere $\rent{\pxthetas : \pxthetasd}$ et suppose que 
$\, \ln \pxtheta$ est d\'eveloppable  
en s\'erie de Taylor. En ne gardant que les termes jusqu'\`a 
l'ordre deux, on obtient 
\beq \label{Fisher1}
\rent{\pxthetas : \pxthetasd} \simeq 
\frac{1}{2} \Delta\thetab^t \Fb \Delta\thetab. 
\eeq
o\`u $\Fb$ est la matrice d'information de Fisher :
\beq \label{Fisher2}
\Fb 
=\esp{\frac{\partial^2}{\partial \thetab^t\partial \thetab} 
\, \ln \pxtheta | _{\thetab=\thetab^*}}.  
\eeq

\subsection{Cas d'un vecteur ou d'un processus al\'eatoire}

Toutes ces d\'efinitions sont facilement \'etendues au cas 
d'un vecteur al\'eatoire ou d'un processus al\'eatoire stationnaire. 
Par exemple, il est facile de montrer que 
l'entropie d'un vecteur al\'eatoire de dimension $n$ 
avec une densit\'e gaussienne ${\cal N}(\bm{0,\Rb})$ est 
\beq
H=\frac{n}{2} \, \ln (2\pi) + \frac{1}{2} \, \ln ( | \det{\Rb} | ) 
\eeq
et que l'entropie relative entre deux lois gaussiennes ${\cal N}(\bm{0,\Rb})$ 
et ${\cal N}(\bm{0,\Sb})$ est
\beq
D=-\frac{1}{2} \left( 
\tr{\Rb\Sb^{-1}}-\log\frac{ | \det{\Rb} | }{ | \det{\Sb} | } - n
\right). 
\eeq
De m\^eme, on montre que pour un processus al\'eatoire stationnaire 
et gaussien dont la matrice de covariance est Toeplitz, on a 
\beq 
\lim_{n\lra\infty} \frac{1}{n} H(p) 
= \frac{1}{2\pi } \, \int_{-\pi}^{\pi} \, \ln S(\omega) \d{\omega}
\eeq
o\`u $S(\omega)$ est sa densit\'e spectrale de puissance (dsp), 
et pour deux processus stationnaires 
et gaussiens de densit\'es spectrales de puissance  
$S_1(\omega)$ et $S_2(\omega)$, on a 
\beq 
\lim_{n\lra\infty} \frac{1}{n} D(p_1 : p_2) 
= \frac{1}{4\pi } \, \int_{-\pi}^{\pi} \left(
\frac{S_1(\omega)}{S_2(\omega)} 
- \, \ln \frac{S_1(\omega)}{S_2(\omega)} -1 
\right) \d{\omega}
\eeq 
et on retrouve la distance de Itakura-Saito 
\cite{Itakura70,Knockaert93,Schroeder84} en analyse spectrale. 

\subsection{Principe du maximum d'entropie (PME)}
 
Lorsqu'on doit attribuer une loi de probabilit\'e \`a une variable $X$ 
sur laquelle on a une information partielle, il est pr\'ef\'erable 
de choisir la loi d'entropie maximale parmi toutes les lois compatibles avec
cette information. 
La loi ainsi choisie est la moins compromettante au sens qu'elle
ne contient que l'information disponible 
(elle n'introduit pas d'information suppl\'ementaire). 

En termes math\'ematiques, consid\'erons la variable $X$ et 
supposons que l'information disponible sur $X$ s'\'ecrit   
\beq \label{contraintes}
\esp{\phi_k(X)}=d_k, \quad k=1,\ldots, K. 
\eeq
o\`u $\phi_k$ sont des fonctions quelconques. 
\'Evidemment, il existe une infinit\'e de lois $p(x)$ qui satisfont ces
contraintes. Alors le PME s'\'ecrit 
\beq \label{PME}
\wh{p}(x) = \argmax{p\in{\cal P}}{\ent{p}=-\int p(x) \, \ln p(x) \d{x}} 
\eeq 
o\`u 
\[
{\cal P}= \left\{p(x) : \int \phi_k(x) p(x) \d{x} = d_k, 
\quad k=0,\ldots, K\right\}
\]
avec $\phi_0=1$ et $d_0=1$ pour la contrainte de normalisation. 

Sachant que $\ent{p}$ est une fonction concave de $p$ et que les contraintes 
(\ref{contraintes}) sont lin\'eaires en $p$, la solution s'\'ecrit 
\beq
\wh{p}(x)=\frac{1}{Z(\lambdab)} \expf{-\sum_{k=1}^K \lambda_k \phi_k(x)}
\eeq
o\`u $Z(\lambdab)$ est la fonction de partition 
$Z(\lambdab)=\disp{\int} \exp[-\sum_{k=1}^K \lambda_k \phi_k(x)] \d{x}$ 
et 
$\lambdab=[\lambda_1,\ldots,\lambda_K]^t$ v\'erifie  
\beq
\frac{1}{Z(\lambdab)} \int \phi_k(x) \expf{-\sum_{k=1}^K \lambda_k \phi_k(x)} 
= d_k, \quad k=1,\ldots, K.
\eeq
La valeur maximale de l'entropie est 
\beq
H_{\max} = \, \ln Z(\lambdab) + \lambdab^t\yb. 
\eeq
Le probl\`eme d'optimisation (\ref{PME}) s'\'etend facilement en 
rempla\c cant l'entropie $H(p)$ par l'entropie relative $D[p : q]$ 
o\`u $q(x)$ est une loi \apriori. 
Pour plus de d\'eveloppements sur ce sujet on peut se r\'ef\'erer \`a  
\cite{Jaynes82,Kullback59,Shore80,Shore81a} et \`a 
\cite{Djafari94,Bercher95a,LeBesnerais93a,Borwein91a}. 

\section{Entropie en s\'eparation de sources}

Le mod\`ele le plus simple en s\'eparation de sources est 
$\xb=\Ab \, \sb$ o\`u, $\sb$ est le vecteur sources, $\xb$ est 
le vecteur des mesures et $\Ab$ est la matrice du m\'elange, 
suppos\'ee inversible en g\'en\'eral. Le probl\`eme est souvent 
pos\'e comme celui de l'estimation d'une matrice de s\'eparation 
$\Bb=\Ab^{-1}$ ou $\Bb=\Sigmab\, \Lambdab\, \Ab^{-1}$.   
$\Sigmab$ est une matrice de permutation d'indices et $\Lambdab$ une 
matrice diagonale, de telle sorte que les composantes du vecteur 
$\yb=\Bb\xb$ soient ind\'ependantes. 
La notion d'entropie est utilis\'ee \`a ce niveau comme un outil 
pour assurer cette ind\'ependance. 
D'une mani\`ere plus g\'en\'erale, consid\'erons un traitement 
de la forme $y_i=g([\Bb\xb]_i)$ o\`u $g$ est une 
fonction monotone et croissante. On a alors 
\beq
p_Y(\yb)=\frac{1}{\left|{\partial \yb}/{\partial \xb}\right|} p_X(\xb)
\lra 
H(\yb) = -\esp{\ln p_Y(\yb)} 
= \esp{\ln \left|{\partial \yb}/{\partial \xb}\right|} - H(\xb).
\eeq
$H(\yb)$ est utilis\'ee comme une mesure de l'ind\'ependance 
des composantes du vecteur $\yb$ et   
on estime alors la matrice de s\'eparation $\Bb$ en maximisant $H(\yb)$ 
par rapport aux \'el\'ements de cette matrice. 
\`A titre de comparaison, on note que l'estimation de $\Bb$ au sens 
du maximum de vraisemblance s'obtient en maximisant 
\beq
V(\Bb) = \sum_i \, \ln p_i\left([\Bb\xb]_i\right) - \log |\det{\Bb}|  
\eeq
lorsque les sources $s_i$ sont suppos\'ees ind\'ependantes avec $p_i(s_i)$ 
connues. 
 

\section{Entropie en mod\'elisation de signaux}

L'identification de l'ordre d'un mod\`ele en traitement du signal est un 
sujet primordial et encore ouvert. Lorsque l'ordre du mod\`ele 
(dimension du vecteur param\`etre $\thetab$) est fix\'e, l'estimation d'une 
valeur optimale (au sens du maximum du vraisemblance, 
du maximum \aposteriori (MAP)  
ou d'autres estimateurs bay\'esiens) est bien \'etablie, 
mais la d\'etermination de l'ordre du mod\`ele est encore mati\`ere 
\`a discussion. 
Parmi les outils utilis\'es, on peut mentionner l'entropie, ou plus exactement 
$\rent{p(\xb | \thetab^*) : p(\xb | \thetab)}$, o\`u $\thetab^*$ repr\'esente le
vrai vecteur des param\`etres de dimension $k^*$ et $\thetab$ le 
vecteur estim\'e de dimension $k \le k^*$. Le fameux crit\`ere d'Akaike 
\cite{Akaike69,Akaike74,Farrier84,Wax85,Wax91} 
utilise ainsi cette quantit\'e pour d\'eterminer l'ordre optimal 
du mod\`ele dans le cadre sp\'ecifique des mod\`eles lin\'eaires  
(en les param\`etres), des lois gaussiennes et de l'estimation au 
sens du MV \cite{Matsuoka86}. 



\section{Entropie en analyse spectrale}

L'entropie est utilis\'ee de multiples fa\c cons en analyse spectrale. 
La pr\'esentation classique de Burg \cite{Burg67} se r\'esume ainsi : 
\\ 
Soit $X(n)$ un processus al\'eatoire centr\'e et stationnaire, 
dont nous disposons d'un nombre fini d'\'echantillons de la 
fonction d'autocorr\'elation 
\beq \label{contraintes2}
r(k)=\esp{X(n) X(n+k)}
=\frac{1}{2\pi} \int_{-\pi}^{\pi} S(\omega) \expf{\j k\omega} \d{\omega},
\quad k=0,\ldots,K. 
\eeq
La question est d'estimer la densit\'e spectrale de puissance  
\[
S(\omega)=\sum_{k=-\infty}^{\infty} r(k) \expf{-\j k \omega} 
\]
de ce processus.  
Consid\'erons maintenant le probl\`eme 
de l'attribution d'une loi de probabilit\'e $p(\xb)$ au vecteur 
$\underline{X}=[X(0),\ldots,X(N-1)]^t$. Utilisant le PME et en remarquant 
que les contraintes (\ref{contraintes2}) sont quadratiques en 
$\underline{X}$, on obtient une loi gaussienne pour $\underline{X}$. 
Pour un processus centr\'e, stationnaire et gaussien, 
lorsque le nombre d'\'echantillons $N\lra \infty$, 
l'expression de l'entropie devient 
\beq \label{Burg}
H = \int_{-\pi}^{\pi} \, \ln S(\omega) \d{\omega}. 
\eeq
On cherche alors \`a maximiser $H$ sous les contraintes (\ref{contraintes2}). 
La solution est bien connue : 
\beq
S(\omega) = \frac{1}{\disp{
\left| \sum_{k=-K}^{K} \lambda_k\expf{\j k\omega}\right| ^2
}}, 
\eeq
o\`u $\lambdab=[\lambda_0,\cdots,\lambda_K]^t$, les multiplicateurs 
de Lagrange associ\'es aux contraintes (\ref{contraintes2}), 
sont ici \'equivalents aux 
coefficients d'une mod\'elisation AR du processus $X(n)$. 
Notons que dans ce cas particulier, il y a une expression analytique pour 
$\lambdab$, ce qui permet de donner une expression analytique directe 
de $S(\omega)$ en fonction des donn\'ees $\{r(k), k=0,\cdots,K\}$ :
\beq
S(\omega) = \frac{\disp{\bm{\delta} \,\bm{\Gamma}^{-1} \bm{\delta}}}{\disp{\bm{e} \,\bm{\Gamma}^{-1} \bm{e}}},
\eeq
o\`u  $\bm{\Gamma}=\hbox{Toeplitz}(r(0),\cdots,r(K))$ 
est la matrice de corr\'elation des donn\'ees et $\bm{\delta}$ et $\bm{e}$ 
sont deux vecteurs d\'efinis par 
$\bm{\delta}=[1,0,\cdots,0]^t$ et 
$\bm{e}=[1,e^{-\j \omega},e^{-\j 2\omega},\cdots,e^{-\j K\omega}]^t$.  
 
Notons que nous avons utilis\'e le PME pour choisir une loi de 
probabilit\'e pour le processus $X(n)$. 
Ainsi la densit\'e spectrale de puissance estim\'ee dans cette
approche correspond \`a la densit\'e spectrale de puissance du 
processus le plus d\'esordonn\'e 
(le plus informatif !) qui soit compatible avec les donn\'ees 
(\ref{contraintes2}). 

Une autre approche consiste \`a maximiser l'entropie relative 
$\rent{p(\xb) : p_0(\xb)}$ ou minimiser la distance de Kullback 
$\kullback{p(\xb) : p_0(\xb)}$ o\`u $p_0(\xb)$ est une loi \apriori 
sous les m\^eme contraintes. Le choix de cette loi est alors primordial. 
\'Evidemment, en choisissant $p_0(\xb)$ uniforme, on retrouve le cas 
pr\'ec\'edent, 
mais si on choisit une loi gaussienne pour $p_0(\xb)$,  
l'expression \`a maximiser devient
\beq 
\rent{p(\xb) : p_0(\xb)}
= \frac{1}{4\pi } \, \int_{-\pi}^{\pi} \left(
\frac{S(\omega)}{S_0(\omega)} 
- \, \ln \frac{S(\omega)}{S_0(\omega)} -1 
\right) \d{\omega}
\eeq 
lorsque $N\mapsto\infty$, et o\`u $S_0(\omega)$ correspond \`a 
la densit\'e spectrale de puissance  
d'un processus de r\'ef\'erence avec la loi $p_0(\xb)$. 

Une autre approche consiste \`a d\'ecomposer le processus $X(n)$ sur 
une base de Fourier \\ 
$\{\cos k\omega t, \,\sin k\omega t\}$ et consid\'erer 
$\omega$ comme une variable al\'eatoire et $S(\omega)$, une fois normalis\'e, 
comme une loi de probabilit\'e. On d\'ecrit alors le probl\`eme de la 
d\'etermination de $S(\omega)$ comme celui de la maximisation de 
\beq \label{Shann}
- \int_{-\pi}^{\pi} S(\omega) \, \ln S(\omega) \d{\omega}
\eeq
sous les contraintes lin\'eaires (\ref{contraintes2}). 
La solution est de la forme 
\beq
S(\omega) = \expf{\sum_{k=-K}^{K} \lambda_k \expf{\j k\omega}}.
\eeq
La densit\'e spectrale de puissance estim\'ee dans cette
approche correspond \`a la densit\'e spectrale de puissance la plus uniforme 
du processus qui est 
compatible avec les donn\'ees (\ref{contraintes2}). 

Une troisi\`eme approche consiste \`a consid\'erer $S(\omega)$ 
(\`a $\omega$ fix\'e) comme la 
moyenne d'une variable al\'eatoire $Z(\omega)$ pour laquelle nous 
supposons disposer d'une loi a priori $\mu(z)$. 
On cherche ensuite la loi $p(z)$ 
qui maximise $D(p(z);\mu(z))$ sous les contraintes (\ref{contraintes2}). 
Une fois $p(z)$ d\'etermin\'ee, on d\'efinit la solution par 
\beq
S(\omega) = \esp{Z(\omega)}=\int Z(\omega) p(z) \d{z}. 
\eeq
Il est alors int\'eressant de voir que l'expression de $S(\omega)$ 
d\'epend du choix de la loi a priori $\mu(z)$ (voir paragraphe \ref{pbinv}). 
Lorsqu'on choisit pour $\mu(z)$ une loi gaussienne (sur $\ER$) on obtient
\beq \label{Gauss}
H = \int_{-\pi}^{\pi} S^2(\omega) \d{\omega}, 
\eeq
alors que si on choisit une loi de Poisson (sur $\ER_+$), on retrouve 
l'expression de l'entropie (\ref{Shann}). Finalement, si on choisit 
une mesure de Lebesgue sur $[0,\infty]$, on obtient l'expression 
de l'entropie (\ref{Burg}). 
Voir aussi : 
\cite{Burg67,Shore81b,Johnson81,McClellan82,Johnson84,Picinbono90,Borwein91a}. 


\section{Entropie pour la r\'esolution des probl\`emes inverses lin\'eaires} 
\label{pbinv}

Lorsqu'on cherche \`a r\'esoudre un probl\`eme inverse lin\'eaire 
num\'eriquement, on est rapidement amen\'e \`a chercher une solution 
$\wh{\xb}$ pour l'\'equation
\beq
\yb = \Ab \xb, 
\eeq
o\`u $\Ab$ est une matrice de dimensions $(M\times N)$, en g\'en\'eral 
singuli\`ere ou tr\`es mal conditionn\'ee. Bien que les cas $M>N$ 
ou $M=N$ aient les m\^emes difficult\'es que le cas $M<N$, nous 
consid\'erons seulement ce deuxi\`eme cas pour plus de 
clart\'e. 
Dans ce cas, \`a l'\'evidence, soit le probl\`eme n'a pas de solution, 
soit il en poss\`ede une infinit\'e. 
Nous nous pla\c{c}erons dans ce dernier cas o\`u 
la question est de choisir une seule solution. 

Parmi les diff\'erentes m\'ethodes, on peut noter l'utilisation de la norme 
$\| \xb\|^2$ pour ce choix --- la solution de norme minimale : 
\beq
\wh{\xb}_{NM} = \argmax{\{\xb \, : \, \yb = \Ab \xb\}}
{\Omega(\xb)=\| \xb\|^2} 
= \Ab^t (\Ab \Ab^t)^{-1} \yb.   
\eeq
Mais, ce choix permettant d'obtenir une  
unique solution \`a ce probl\`eme n'est pas le seul possible. 
En effet, tout crit\`ere $\Omega(\xb)$ qui est convexe en $\xb$ 
peut \^etre utilis\'e. On peut mentionner en particulier 
\beq
\Omega(\xb)= - \sum_j x_j \, \ln x_j
\eeq
lorsque les $x_j$ sont positifs et lorsque $\sum x_j=1$,  
ce qui, par analogie avec la d\'efinition de l'entropie, assimile 
les $x_j$ \`a  
une distribution de probabilit\'e $x_j=P(U=u_j)$.  
La variable al\'eatoire $U$ peut ou non avoir une r\'ealit\'e physique. 
$\Omega(\xb)$ est alors l'entropie associ\'ee \`a cette variable. 


Une autre approche consiste \`a supposer $x_j=\esp{U_j}$ ou 
encore $\xb=\esp{\Ub}$ o\`u $\Ub$ est un vecteur al\'eatoire, 
qui peut ou non avoir une r\'ealit\'e physique. 
Supposons maintenant que $\Ub$ admet une loi de probabilit\'e 
$\wh{p}(\ub)$ que l'on cherche \`a d\'eterminer. 
En notant que les donn\'ees $\yb=\Ab\xb=\Ab\esp{\Ub}=\esp{\Ab\Ub}$ 
peuvent \^etre consid\'er\'ees comme des contraintes lin\'eaires sur 
cette loi, on peut utiliser de nouveau l'entropie pour d\'eterminer la 
loi $\wh{p}(\ub)$ : 

\beq \label{lag1}
\wh{p}(\ub) 
= \argmax{\{\xb \, : \, \yb=\int \Ab \ub \, p(\ub)\d{\ub}\}}
{D[p(\ub) : \mu(\ub)]}
\eeq
o\`u $\mu(\ub)$ est une loi \apriori dont nous montrerons par la suite 
l'importance. 
La solution est bien connue :
\beq \label{sol1}
\wh{p}(\ub)=\frac{1}{Z(\lambdab)} \mu(\ub) \expf{-\lambdab^t\Ab\ub} 
\eeq
mais le plus int\'eressant est de voir ce que devient $\wh{\xb}=\esp{\Ub}$. 
Bien \'evidemment, $\wh{\xb}$ d\'epend de $\mu(\ub)$. Le tableau qui suit 
donne quelques exemples :
\[
\barr{||l|l|l||} \hline \hline
 \mu(\ub)\propto \exp[-\frac{1}{2}\sum_j u_j^2]
& \wh{\xb} = \Ab^t\lambdab 
& \Ab\Ab^t\lambdab = \yb 
\\ \hline 
\mu(\ub)\propto \exp[-\sum_j  | u_j | ]
& \wh{\xb} = \bm{1} ./ (\Ab^t\lambdab \pm \bm{1})
& \Ab \xb = \yb 
\\ \hline 
\mu(\ub)\propto \exp[-\sum_j u_j^{\alpha-1}\expf{-\beta u_j}], 
\quad u_j>0 
& \wh{\xb} = \alpha \bm{1} ./ (\Ab^t\lambdab + \beta\bm{1})
& \Ab \xb = \yb \\ \hline\hline
\earr
\]
Dans le cas plus g\'en\'eral, rempla\c cant (\ref{sol1}) dans (\ref{lag1}) 
et d\'efinissant 
 $Z(\lambdab) = \disp{\intg} \mu(\ub) \expf{-\lambdab^t\Ab\ub} \d{\ub}$, 
$G(\sb) = \, \ln \disp{\intg} \mu(\ub) \expf{-\sb^t\ub} \d{\ub}$ 
et sa convexe conjug\'ee $F(\xb) = \sup_{\sb}\left\{\xb^t\sb-G(\sb)\right\}$, 
on peut montrer que $\wh{\xb}=\esp{\Ub}$ peut 
\^etre obtenu, soit comme une fonction de son vecteur dual $\wh{\lambdab}$ par 
$\wh{\xb}=G'(\Ab^t\wh{\lambdab})$ o\`u $\wh{\lambdab}$ est solution 
du probl\`eme d'optimisation 
\beq 
\wh{\lambdab}=\argmin{\lambdab}{D(\lambdab)=\, \ln Z(\lambdab)+\lambdab^t\yb}, 
\eeq
soit directement comme la solution du probl\`eme d'optimisation sous
contraintes 
\beq 
\wh{\xb}=\argmin{\{\xb \, : \, \Ab\xb=\yb\}}{F(\xb)}. 
\eeq
$D(\lambdab)$ est appel\'e ``crit\`ere dual'' et $F(\xb)$ ``crit\`ere
primal''. Parfois, il est plus facile de r\'esoudre le probl\`eme dual, 
mais il n'est pas toujours possible d'obtenir une expression explicite 
pour $G(\sb)$ et son gradient $G'(\sb)$. Les fonctions $F(\xb)$ et 
$G(\sb)$ sont convexes conjugu\'ees. 


\section{Conclusions}
La notion d'entropie, vue comme une mesure de la quantit\'e d'information 
dans les r\'ealisations d'une variable al\'eatoire est 
utilis\'ee de multiples fa\c{c}ons dans diff\'erents domaines du traitement 
de l'information. Lors de son utilisation, il est tr\`es important de 
bien pr\'eciser quelle est la variable consid\'er\'ee, quelles sont 
les donn\'ees, quelle est la relation entre les donn\'ees et cette 
variable, et finalement, 
quelle est le crit\`ere optimis\'e. Par exemple, en estimation spectrale, 
nous avons vu comment le choix de la variable al\'eatoire ($X(n)$, 
$S(\omega)$ ou $Z(\omega)$), le choix du crit\`ere 
(entropie ou entropie relative) et le choix de la loi \apriori dans 
le cas de l'entropie relative, peuvent influencer l'expression de la solution. 
Bien entendu, nous n'avons pas discut\'e ici le probl\`eme 
de l'estimation des coefficients de corr\'elation \`a partir 
des \'echantillons du signal. 
Par ailleurs, l'estimation de la densit\'e spectrale de puissance 
d'un processus \`a partir d'une connaissance partielle de ses 
coefficients de corr\'elation 
n'est qu'un cas particulier des probl\`emes inverses lin\'eaires. 


\section*{Remerciements}
L'auteur remercie Odile Macchi et Charles Soussen pour la relecture 
attentive de cet article. 

\section*{Biographie de l'auteur}
Ali Mohammad-Djafari est n\'e en Iran en 1952. 
Il est Ing\'enieur de l'\'Ecole Polytechnique de T\'eh\'eran (1975), 
Ing\'enieur de l'\'Ecole Sup\'erieure d'\'Electricit\'e (1977), 
Docteur-Ing\'enieur (1981) et Docteur-\`es-Sciences Physiques (1987) 
de l'Universit\'e de Paris-Sud, Orsay. 
Il travaille depuis 1977 au Laboratoire des Signaux et Syst\`emes 
au sein du groupe 
``Probl\`emes Inverses en Traitement du Signal et Imagerie''. 
Charg\'e de Recherche au CNRS depuis 1983, il s'int\'eresse \`a la     
r\'esolution des probl\`emes inverses en utilisant des 
m\'ethodes d'inf\'erence probabilistes. 
Parmi les applications de ses th\`emes de recherche on peut mentionner :  
restauration et reconstruction des signaux mono- ou multi- variables,
imagerie tomographique \`a rayons X, \`a ondes diffract\'ees ou 
par courants de Foucault en contr\^ole non destructif (CND).  

{\small


\def\ieeeSSC{IEEE Transactions on Systems Science and Cybernetics}
\def\ieeeSP{IEEE Transactions on Signal Processing}			
\def\ieeeP{Proceedings of the IEEE}					
\def\ieeeIT{IEEE Transactions on Information Theory}			
\def\TS{Traitement du Signal}						
\def\siamCO{SIAM Journal of Control}					
\def\AISM{Annals of Institute of Statistical Mathematics}		
\def\ieeeASSP{IEEE Transactions on Acoustics Speech and Signal Processing}
\def\ieeeAC{IEEE Transactions on Automatic and Control}			

\def\jan{janvier\xspace}
\def\feb{f\'evrier\xspace}
\def\mar{mars\xspace}
\def\apr{avril\xspace}
\def\may{mai\xspace}
\def\jun{juin\xspace}
\def\jul{juillet\xspace}
\def\aug{ao\^ut\xspace}
\def\sep{septembre\xspace}
\def\oct{octobre\xspace}
\def\nov{novembre\xspace}
\def\dec{d\'ecembre\xspace}
\def\Jan{January\xspace}	
\def\Feb{February\xspace}
\def\Mar{March\xspace}
\def\Apr{April\xspace}
\def\May{May\xspace}
\def\Jun{June\xspace}
\def\Jul{July\xspace}
\def\Aug{August\xspace}
\def\Sep{September\xspace}
\def\Oct{October\xspace}
\def\Nov{November\xspace}
\def\Dec{December\xspace}

}

\end{document}